\documentclass[11pt]{article}
\usepackage{geometry}                
\usepackage[parfill]{parskip}    
\usepackage{subcaption}
\usepackage{graphicx}
\usepackage{caption}

\usepackage{amsmath, amsthm, amsfonts}
\usepackage{amssymb}

\usepackage{tikz-cd}
\usetikzlibrary{positioning}
\usepackage{caption}
 
\newcommand{\sect}[1]{\setcounter{equation}{0}\section{#1}}

\usepackage{color}
     \usepackage[colorlinks]{hyperref}
\hypersetup{linkcolor=blue,%
citecolor=red,%
urlcolor=cyan}
\usepackage{url} 

\usepackage{epstopdf}
\begin{document}

\title{SUSY hierarchies of 
Jaynes-Cummings Hamiltonians \\
with different detuning parameters
}

\author{
\.{I}smail Burak Ate\c s$^a$\footnote{ismailburakates@gmail.com, ORCID: \href{http://orcid.org/0000-0001-8262-2572}{0000-0001-8262-2572}}, 
\c Seng\"ul Kuru$^a$\footnote{sengul.kuru@science.ankara.edu.tr-Corresponding author, ORCID: \href{http://orcid.org/0000-0001-6380-280X}{0000-0001-6380-280X}}, 
Javier Negro$^b$\footnote{jnegro@uva.es, 
ORCID: \href{http://orcid.org/0000-0002-0847-6420}{0000-0002-0847-6420}}
\\ 
\\
$^a$Department of Physics, Faculty of Science, Ankara University,\\ 06100 Ankara, T\"urkiye\\
$^b$Departamento de F\'i{}sica Te\'orica, At\'omica y \'Optica and IMUVA, \\Universidad de Valladolid, 47011 
Valladolid, Spain\\
}

\maketitle
	
\begin{abstract}
The aim of this work is to show how  supersymmetric (SUSY) quantum mechanics can be applied to the Jaynes-Cummings (JC) Hamiltonian of quantum optics. These SUSY transformations connect pairs of Jaynes-Cummings Hamiltonians characterized by different detuning parameters  as well as 
 Jaynes-Cummings to  anti-Jaynes-Cummings Hamiltonians. Therefore, JC Hamiltonians can be classified in hierarchies or sequences which are connected through SUSY transformations. As a byproduct, the symmetries of JC Hamiltonians are found as well as the special case of a sequence of resonant-like interacting systems having  the form of a simple shape invariant JC Hamiltonian hierarchy.
\end{abstract}

\sect{Introduction}
The Jaynes-Cummings  Hamiltonian is a simple theoretical model that describes the  quantum  interaction of a two-level atom and radiation (electromagnetic field quanta) \cite{jaynes63}. It has given rise to a huge number of applications to multiple configurations of the interaction of matter and radiation \cite{rempe87,gerry04}. 
The JC Hamiltonian is an approximation in the Rotating Wave Approximation (RWA) of the quantum Rabi model of quantum optics. This approximation is good for low energy states and when the detuning (difference of atomic and radiation frequencies) is small with respect to the coupling radiation-atom parameter. In some coupling regimes the counter-rotating terms must be considered (for more details see Refs. \cite{Rossatto17,Larson24} for
spectral classification of coupling regimes in the quantum Rabi model and \cite{Solano19,Huang20} for ultrastrong coupling regimes).

In this work we will consider the JC Hamiltonian corresponding to the Rabi quantum model in the RWA without any limitations on the parameters. This Hamiltonian may also be useful in condensed matter and solid-state physics. Thus, when the JC Hamiltonian is applied in quantum optics or in any other problem one should be aware of the corresponding restrictions on the parameters and energy values. Along this work we will deal with properties of the JC model which do not depend on these restrictions.

The wide interest of the JC model is based on the simplicity, as well as to its complete solvability in the RWA,  when the frequency of the applied electromagnetic field is close to the atomic transition frequency of the atom (resonance regime) so that the high-frequency counter-rotating terms in the Hamiltonian are neglected. This model explains the quantum character
\cite{Haroche96} of properties such as purely quantum vacuum field effects, Rabi oscillations or quantum collapses and revivals \cite{Chong25} of some expected value evolutions.

The essentials of supersymmetric  quantum mechanics methods, often referred to as Darboux \cite{matveev91}, intertwining or factorization \cite{infeld51}, depending on the context, can be found  in many reviews and  textbooks 
\cite{fernandez10,cooper95,junker96}.  The intertwining method can be applied differential,  integral or matrix forms of Hamiltonians in quantum mechanics (or in mathematics) in order to provide a unified approach to constructing exactly solvable linear and nonlinear problems.
It is also a way to obtain a whole hierarchy (a sequence) of exactly solvable Hamiltonians, starting from just one solvable system. The eigenvalues and eigenfunctions of the Hamiltonian operators in each hierarchy are connected by means of intertwining  operators.  These methods have been applied to  a  variety of scalar Hamiltonians, but here their relevance relies in the fact that they have also been extended to deal with matrix Hamiltonians. For example, in \cite{negro04} the ingiedients of Darboux transformations for the JC Hamiltonian are given, in \cite{hussin06}  generalized JC Hamiltonians are considered by shape-invariant hierarchies and their SUSY partners and in \cite{kuru21} Dirac-like Hamiltonians associated to Schr\"odinger factorizations are discussed. Thus, our aim in this work consists in translating SUSY techniques to the matrix JC model in a new original way.

There is already a considerable list of references where supersymmetry tools have been applied to study different aspects of the JC model. We will mention a number of representative references not exhaustive at all. For instance, Ref.~\cite{negro24} was devoted to the SUSY relation of JC and anti-JC Hamiltonians; in Refs.~\cite{lara24,lara05} Jaynes-Cummings and anti-Jaynes-Cummings models (which include only the non-conserving interaction terms) were revisited by means of  Lie algebras and superalgebras; differential realizations of superalgebras were applied in Refs.~\cite{alhaidari06} and \cite{panahi15}. Refs.~\cite{lara13,lara21} proposed an extension that includes nonlinear processes with qubits exchanges. Stark shift and nonlinear Kerr-like medium have also been considered, for instance in \cite{mubark23}. Other new similar systems have been introduced, for example
\cite{alexio07} defines a system  coupling  a two-level atom with a two-dimensional supersymmetric system involving two shape invariant potentials.
The coupling of two isospectral JC Hamiltonians is introduced in \cite{castanos13}; applications to quantum information technologies are frequent, for instance see \cite{solano03,ismail25}.


In this work, we want to present a novel approach of SUSY techniques to the JC model, which is essentially different to all of these previous references, up to our knowledge.  
We will apply in a systematic way the techniques presented in \cite{negro04,hussin06} to the JC Hamiltonian. We find new intertwining operators depending on creation/annihilation operators made of atomic $\sigma^\pm$ and radiation field $a^\pm$ basic operators which are well defined in the quantum Hilbert space.  These intertwining operators supply us with the relevant symmetries. We will also find original shape-invariance properties of the JC hierarchies. Thus, the main results presented here are completely new.

As a first step, we write the JC Hamiltonian in a suitable matrix form. Then, we will arrive to a hierarchy of JC Hamiltonians characterized  by particular values of detuning parameters depending on the
 difference of atomic transition and field frequencies. The JC Hamiltonians in each hierarchy are connected by triangular intertwining operators, which are different from those applied earlier in quantum optics. This means that the Hamiltonians in each sequence are quite similar (almost isospectral) and must share very similar properties. In addition, this procedure will allow us to obtain symmetries of JC Hamiltonians, such as the excitation number, in a natural way.

The scheme of the paper is as follows. In Section~2 we will discuss the form of the JC Hamiltonian adequate to the application of SUSY transformations. We consider physical as well as nonphysical states necessary for a complete understanding of SUSY transformations. Section 3 is devoted to a review of the ingredients of SUSY transformations that will be used later in the JC matrix context. Sections 4 and 5 show the results corresponding to several examples. Some of them are related to true JC Hamiltonians while others to the so called  anti-JC Hamiltonians. Section 6 addresses the special case of a resonant shape invariant Hamiltonian hierarchy. This particular case is similar to other examples known for the two dimensional Dirac-Weyl Hamiltonians found in the continuous tight-binding approximation of  graphene \cite{kuru21}. In this way, we show a connection between two  areas far apart, quantum optics and condensed matter, where Hamiltonians with a similar structure are applied.

\sect{Matrix form of Jaynes-Cummings Hamiltonian}
The Jaynes-Cummings Hamiltonian (in the  RWA) has the following form \cite{gerry04}:
\begin{equation}\label{H1}
H_{JC}=H_A+H_F+H_{AF}=\frac{\hbar \omega _0}2 \sigma _z+\hbar \omega a^+a^-\sigma_0+\hbar \mu \left(\sigma _{-}a^++\sigma _{+}{a^-}\right)
\end{equation}
where $H_A$ is the Hamiltonian corresponding to a free two-level  atom of energy difference $\hbar \omega _0$, $H_F$ is for the field and  $H_{AF}$ for the interaction between field and atom in the electric dipole approximation \cite{jaynes63, gerry04}.  Here,  $a^\pm$ are photon creation/annihilation operators of frequency $\omega$, while the usual   Pauli matrices ($\sigma_z$, $\sigma^\pm$ and the identity $\sigma_0$) describe the atomic part. The coupling coefficient $\mu$ depends on the interaction with the field. 

Along this work we will consider the Hilbert space of the tensor product of the two-dimensional atomic space generated by the ground and excited states 
$|g\rangle$, $|e\rangle$, respectively, and the radiation quantum space of frequency $\omega$, generated by the number states $|n\rangle$ (sometimes we
write $\psi_n$ instead of $|n\rangle$,  when we use differential realizations).
We make use of the standard notation:
\[
|g\rangle \otimes |n\rangle :=  \left(\begin{array}{c}
0
\\
|n\rangle
\end{array}\right)
:=  
\left(\begin{array}{c}
0
\\
\psi_n
\end{array}\right)
,\quad
|e\rangle \otimes |n\rangle := \left(\begin{array}{c}
|n\rangle
\\
0
\end{array}\right) :=  \left(\begin{array}{c}
\psi_n
\\
0
\end{array}\right)
\]
For the operators a similar notation is applied, for instance
\[
\sigma_x\otimes a^+ :=  \left(\begin{array}{cc}
0 & a^+
\\
a^+ & 0
\end{array}\right)
\]
For all the systems of this work, the underlying  Hilbert space of any JC Hamiltonian is the same and the intertwining operators between JC systems connect two isomorphic Hilbert spaces.

Let us  introduce a new interacting parameter  
$\beta =\mathit{\hbar \mu}$ and a detuning parameter $\alpha =\hbar (\omega _0-\omega )$  of the difference of  atomic transition frequency $\omega_0$ and     field frequency   $\omega$.
Then, we have the following matrix form for the Hamiltonian (\ref{H1}):
\begin{equation}\label{H2m}
H_{\mathit{JC}}=\left(\begin{array}{cc}\hbar \omega a^+a^-+\frac{\hbar \omega_0}{2}
&\beta a^-
\\[1.ex]
\beta a^+&\hbar \omega a^+a^--\frac{\hbar \omega_0}{2}
\end{array}\right)
\end{equation}
If we use the commutator $[a^-,a^+]=1$ in the first element of the above matrix,  
we can rewrite the Hamiltonian $H_{\mathit{JC}}$ in the special form
\begin{equation}\label{H3m}
H_{\mathit{JC}}=
\left(\begin{array}{cc}\hbar \omega a^- a^{+}+\frac{\alpha}{2}
&\beta a^-
\\[1.ex]
\beta a^{+}&\hbar \omega a^{+}a^--\frac{\alpha}{2}
 \end{array}\right) - \frac{\hbar\omega}2 \sigma_0
\end{equation}
where 
$\sigma_0$ is the unit matrix.  Since the constant term on the r.h.s.~has no effect on the dynamics,  it can be neglected.


\subsection{Physical and nonphysical solutions of the HO}
Hereafter, the above JC Hamiltonian will be re-expressed as $H_{\rm JC}$ in the following final notation
\begin{equation}\label{H3m}
H_{JC}=
\hbar \omega
\left(
\begin{array}{cc}  
a^- a^+ +\delta &\lambda a^-
\\[1ex]
\lambda a^+ & a^+a^- -\delta
\end{array}\right)
:= \hbar \omega H_{\rm JC}
\end{equation}
where 
\begin{equation}\label{param}
\lambda = \frac{\beta}{\hbar\omega}\,,\qquad \delta = \frac{\alpha}{2\hbar\omega} =
\frac{\omega_0-\omega}{2\omega}
\end{equation} 
We will always deal with the Hamiltonian 
$H_{\rm JC}$, where we use the detuning parameter $\delta$  as well as the interacting coefficient $\lambda$
in $\hbar \omega$-units as indicated in (\ref{param}). 
The creation/annihilation operators $a^\pm$ of the HO satisfy the standard commutation
\begin{equation}
[a^-,a^+] = 1
\end{equation}
We will make use of a the following differential realization for these operators,
\begin{equation}\label{aas}
a^\pm = \frac{1}{\sqrt{2}}\, \left(\mp\partial_x + x\right)\,
\end{equation}

The physical eigenfunctions $\psi_n:=\psi_n(x)$ of the HO in this realization are defined in the usual way.  These eigenfunctions for the number states (the energy eigenstates) of one-dimensional harmonic oscillator can be found in any standard quantum mechanics textbook. We will write them here  explicitly to fix the notation:
\begin{equation}
\begin{array}{ll}
a^-\psi_0= 0 \ \implies\ &\psi_0(x)= \frac1{\pi^{1/4}} e^{-x^2/2}
\\[2ex]
a^+\psi_{n-1} = \sqrt{n}\,\psi_{n},\quad
&a^-\psi_n = \sqrt{n}\,\psi_{n-1}\,,
\\[2.ex]
a^+a^-\psi_n = N \psi_n= n \psi_n,\quad &n=0,1,2,3,...
\end{array}
\end{equation}
In ket notation, this is
\begin{equation}
|0\rangle \ 
\xrightarrow{\ a^+\ } |1\rangle \dots
\  
\xrightarrow{\ a^+\ } |n-1\rangle
\xrightarrow{\ a^+\ } 
\  |n\rangle\ \dots
\end{equation}
where $N=a^+a^-$ is the number operator and $|n\rangle$ represents the state of $\psi_n$. We say that $|0\rangle$ is the physical ground state annihilated by $a^-$. We are also taking $\hbar \omega$-units for the eigenvalues of $N$, consistent the previous  
Hamiltonian $H_{\rm JC}$.  

In future sections, we will  make use of the following set of nonphysical  eigenfunctions (they are not square integrable) of the HO,
\begin{equation}
\begin{array}{ll}
a^+\phi_{-1} =0 \ \implies \ &\phi_{-1}(x):= \frac1{\pi^{1/4}} e^{x^2/2}  
\\[2.ex]
a^-\phi_{-n}:=\sqrt{-n}\,\phi_{-n-1} ,\qquad   &a^+\phi_{-n-1}:={\sqrt{-n}}\,\phi_{-n}
\\[2.ex]
a^+a^-\phi_{-n} = N \phi_{-n}= -n \phi_{-n},\quad &n= 1,2, 3\dots
\end{array}
\end{equation}
\begin{equation}
\dots \ |-n-1\rangle_{\rm np} \ 
\xleftarrow{\ a^-\ }
\  |-n\rangle_{\rm np}
\xleftarrow{\ a^-\ }
\ \dots \ 
|-2\rangle_{\rm np}
\xleftarrow{\ a^-\ }
\  |-1\rangle_{\rm np}
\end{equation}
In this case, $|-1\rangle_{\rm np}$ is the ket for the  ``highest'' nonphysical state annihilated by $a^+$.
There is a simple relation between physical and nonphysical eigenfunctions given by
\[
\psi_n(i x) = i^n \phi_{-1-n}(x)\,,\qquad n=0,1,2,3\dots 
\]

\subsection{Eigenfunctions and eigenvalues of JC Hamiltonian}

\begin{figure}[h!]
\begin{center}
\includegraphics[width=0.50\textwidth]{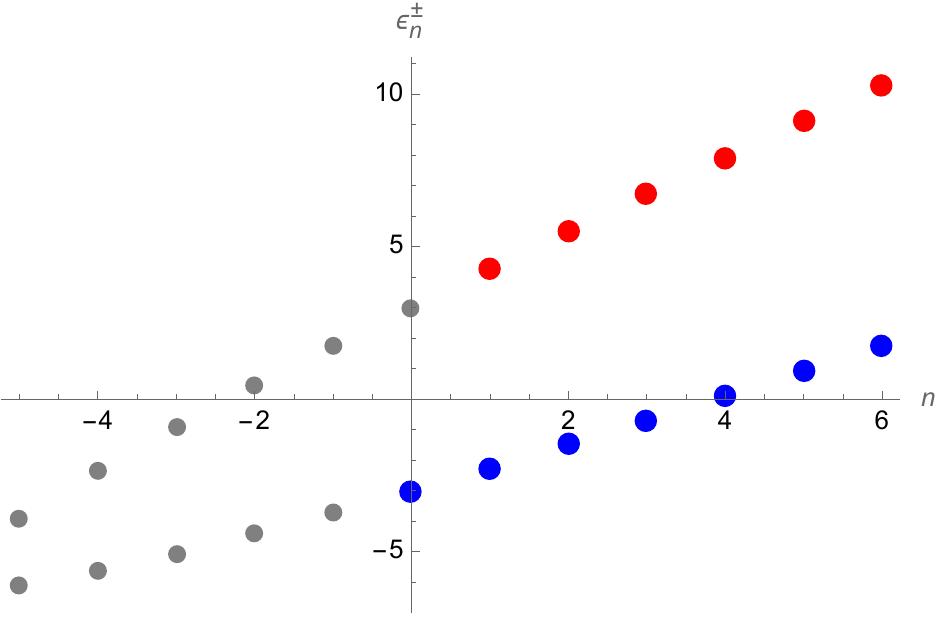}
\end{center}
\caption{\small Eigenvalues of JC for $\delta=3$, $\lambda=1.25$. Blue
(for $\varepsilon_n^-$) and red (for $\varepsilon_n^+$) points represent the (infinite) physical spectrum (\ref{phys}). Points in
gray are for a (finite) nonphysical spectrum (\ref{nphys}).  
\label{f1}
}
\end{figure}

It is convenient, at this stage, to make use of  wave functions for the photon states. Later we will return to the states formulation. As usual, we can find the physical eigenstates  (here they are spinorial eigenfunctions) of the JC Hamiltonian inside
the subspace \cite{gerry04}
\begin{equation}\label{vn}
V_n =\left(\begin{array}{l}
c_1 \psi_{n-1}
\\[1ex]
c_2\psi_n
\end{array}\right)\,,
\qquad c_1,c_2\in \mathbb C
\end{equation}
We will discuss later the connection of $V_n$ with  intertwining operators.
Then, the eigenvalues and (not normalized) eigenfunctions are as follows ($\varepsilon_n= E_n/\hbar\omega$)
\begin{equation}\label{phys}
\begin{array}{lll}
\varepsilon_n^\pm = n \pm \sqrt{\delta^2 +n \lambda^2},\quad 
&\Psi_n^\pm = \left(\begin{array}{c}
(\delta\pm \sqrt{\delta^2 +n \lambda^2})\,\psi_{n-1}
\\[1ex]
\sqrt{n}\,\lambda\, \psi_n
\end{array}\right),\quad &n=1,2,\dots
\\[4.5ex]
\varepsilon^-_0 = -\delta,\qquad
&\Psi^-_0 = \left(\begin{array}{c}
0
\\[1ex]
 \psi_0
\end{array}\right),\quad &n=0
\end{array} 
\end{equation}
For the values $n=1,2,\dots$ there are two types of excited solutions $\Psi_n^\pm$, while only one ground single state, $\Psi^-_0$, is obtained for $n=0$ (see Fig~\ref{f1}).

Next, we will consider some nonphysical eigenfunctions of the JC Hamiltonian.
In order to find them, we make use of the previous nonphysical solutions of the HO.
We  construct this kind of  formal eigenfunctions of the JC Hamiltonian inside
the subspace 
\begin{equation}
{\cal W}_{-n} =\left(\begin{array}{l}
c_1 \phi_{-n-1}
\\[1ex]
c_2\phi_{-n}
\end{array}\right)\,,
\qquad c_1,c_2\in \mathbb C
\end{equation}
where $\phi_{-n}$ are the eigenfunctions defined above. These kind of solutions are as follows
\begin{equation}\label{nphys}
\begin{array}{ll}
\epsilon_{-n}^\pm = -n \pm \sqrt{\delta^2 -n \lambda^2},\qquad
&\Phi_{-n}^\pm = 
\left(\begin{array}{c}
(\delta\pm \sqrt{\delta^2 -n \lambda^2})\,\phi_{-n-1}
\\[1ex]
\sqrt{-n}\lambda\, \phi_{-n}
\end{array}\right),\quad n= 1, 2,\dots
\\[3.5ex]
\epsilon^+_{0} = \delta,\qquad
&\Phi^+_{0} = \left(\begin{array}{c}
\phi_{-1}
\\[1ex]
0
\end{array}\right),\quad n=0
\end{array}
\end{equation}
Note that the energies $\epsilon_{-n}^\pm$ of the nonphysical solutions are real only for a finite number of $n$-values due to the condition $\delta^2 -n \lambda^2\geq 0$.  These nonphysical eigenfunctions with real eigenvalues will also  be useful to get physical partner JC Hamiltonians.
The physical and nonphysical eigenvalues above mentioned have been plotted in Fig~\ref{f1}.

\sect{Construction of JC Hamiltonian partners}

If we make use of the differential realization (\ref{aas}), the JC Hamiltonian has the following expression
\begin{equation}
H_{\rm JC}^{(0)}:=\left(
\begin{array}{cc}  
a^- a^++\delta &\lambda a^-
\\[1.5ex]
\lambda a^+ & a^+a^- -\delta
\end{array}\right)
=
\left(\begin{array}{cc}  
\frac12(-\partial_x^2 +x^2+1)+\delta & \frac{\lambda}{\sqrt{2}}(\partial_x+x)
\\[1.5ex]
 \frac{\lambda}{\sqrt{2}}(-\partial_x+x) & \frac12(-\partial_x^2 +x^2-1)-\delta
\end{array}\right)
\end{equation}
which can be rewritten as
\begin{equation}\label{difH}
H_{\rm JC}=
-\frac12\partial_x^2 + V(x) +
\frac{\lambda}{\sqrt{2}}\,\gamma\, \partial_x 
\end{equation}

where
\begin{equation}
V(x) = \left(\begin{array}{cc}   
\frac12(x^2+1)+\delta & \frac{\lambda}{\sqrt{2}}x
\\[1.5ex]
 \frac{\lambda}{\sqrt{2}}x & \frac12( x^2-1)-\delta
\end{array}\right),
\qquad
\gamma = 
\left(
\begin{array}{cc}  
0 & 1
\\[1.5ex]
-1 & 0
\end{array}\right)
\end{equation}
Then, in this context, we define the partner  Hamiltonians $\widetilde H_{\rm JC}$ of the JC Hamiltonian
$ H_{\rm JC}$ by the following two conditions.
\begin{itemize}
\item[(i)] The differential realization of $\widetilde H_{\rm JC}$  has the same structure 
(\ref{difH}):
\begin{equation}
\widetilde H_{\rm JC}=
-\frac12\partial_x^2 + \widetilde V(x) +
\frac{\tilde \lambda}{\sqrt{2}} \gamma \partial_x
\end{equation} 
where $\widetilde V(x)$  has the same functional form as $V(x)$ but possibly with different parameters, $\tilde \delta, \tilde \lambda$. This means that they satisfy a shape invariance condition for these matrix Hamiltonians.
\item[(ii)] There is an intertwining differential operator
\begin{equation}
L = \frac1{\sqrt2}\big(\partial_x -W(x)\big)
\end{equation}
where $W(x)$ is a matrix,
connecting both of them
\begin{equation}
L\, H_{\rm JC}= \widetilde H_{\rm JC}\, L
\end{equation}
\end{itemize}

It can be shown that the  $W$ matrix can be obtained from two (linearly independent) eigenfunctions
$\Psi_{\rm s1}$ and $\Psi_{\rm s2}$, called seed functions, which may be physical or  nonphysical, of the initial JC Hamiltonian,
\begin{equation}
H_{\rm JC}\Psi_{\rm si}= \varepsilon_i \Psi_{\rm si}\,,\qquad i=1,2
\end{equation}
In a first step, we define an auxiliary matrix $M$ obtained by arranging these two solutions as two columns $M(x)= (\Psi_{\rm s 1},\Psi_{\rm s2})$, such that
$L M(x) =0$. Then, $W$ is given by 
\begin{equation}
W(x) = M_x(x) M^{-1}(x)
\end{equation}
where $M_x(x)$ is for the derivative $d\, M(x)/d\,x$. One can show that the new Hamiltonian $\widetilde H(x)$ has a new potential
$\widetilde V(x)$ given by
\begin{equation}
\widetilde V(x) =  V(x) + \Delta V(x)
\end{equation}
with
\begin{equation}
\Delta V(x) = - W_x(x)   -  \frac{\lambda}{\sqrt{2}}\, [\, W(x) ,\gamma\,]
\end{equation}

In the following, we will apply this program to find partners of $H_{\rm JC}$ by choosing different
intertwining operators $L$, or equivalently, by choosing different pairs of seed
solutions. The details can be found in \cite{negro04,hussin06}.


\sect{First example: anti-JC Hamiltonians}

\begin{figure}[h!]
\begin{center}
\includegraphics[width=0.40\textwidth]{figure1.pdf}
\quad
\includegraphics[width=0.40\textwidth]{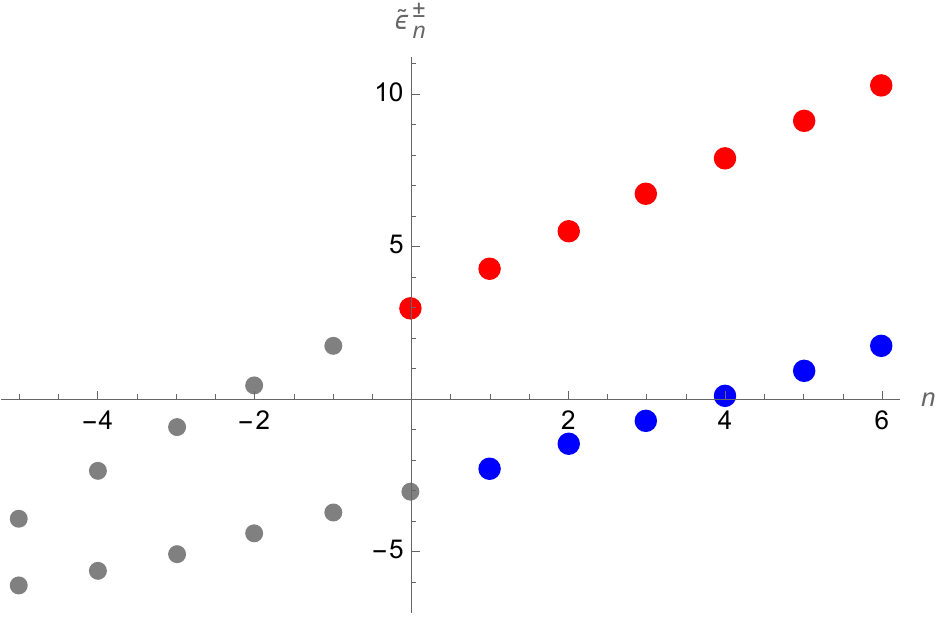}
\end{center}
\caption{\small The eigenvalues of the initial JC Hamiltonian (left) and those of anti-JC (right) for $\delta=3$, $\lambda=1.25$. Blue and red points are for the physical spectrum, $\varepsilon_n^-$ and  $\varepsilon_n^+$, respectively, while gray are for nonphysical spectrum. Notice the difference  of the ground states energies with respect to the initial JC Hamiltonian (left graphic) as mentioned after (\ref{l00}). 
\label{f2}}

\end{figure}

In this initial case, we choose the seed states $\Psi_{\rm s1}$ and $\Psi_{\rm s2}$  as follows 
\begin{equation}
\Psi_{\rm s1}\ \to \ \Psi^-_0,\qquad 
\Psi_{\rm s2}\ \to \ \Phi^+_0
\end{equation}
In other words, one of them is the physical and the other is the nonphysical ground states, as defined in (\ref{phys}) and (\ref{nphys}). Then, we obtain the following results (now we express them in terms of operators, leaving the differential realization): 

(i) The intertwining operator has a diagonal form given by
\begin{equation}
L= \left(\begin{array}{cc}
-a^+ & 0
\\
0 & a^-
\end{array}\right)
\end{equation}
(ii) The partner Hamiltonian will be
\begin{equation}\label{ajc}
\widetilde H_{\rm JC}= H_{\rm aJC}=\left(
\begin{array}{cc}  
a^+ a^- + \delta &- \lambda a^+
\\[1ex]
-\lambda a^- & a^-a^+ -\delta
\end{array}\right)
\end{equation}
This partner Hamiltonian is a kind of  anti-JC Hamiltonian $H_{\rm aJC}$ \cite{negro24,lara05,solano03}. The aJC Hamiltonian is obtained when only the high frequency terms of the total $H_{AF}$ are conserved. We can write this partner Hamiltonian in different equivalent forms by means of a unitary matrix
$R$ such that  
\begin{equation} \label{tt}
\widetilde{H}^R_{\rm JC}= R\widetilde{H}_{\rm JC} R^{-1}
\end{equation}
Let us consider just the following cases for $R$:
\begin{equation} \label{t}
T= \sigma_z
\quad \to \quad 
\widetilde{H}^T_{\rm JC}=
\left(\begin{array}{cc}  
a^+ a^- + \delta & \lambda a^+
\\[1ex]
\lambda a^- & a^-a^+ -\delta
\end{array}\right)
\end{equation}
\begin{equation} \label{s}
 S=  \sigma_y
\quad \to \quad 
\widetilde{H}_{\rm aJC}^{(0)}:=\widetilde{H}^S_{\rm JC}=
\left(\begin{array}{cc}  
a^- a^+ - \delta & \lambda a^-
\\[1ex]
\lambda a^+ & a^+a^- +\delta
\end{array}\right) = H_{\rm JC}(-\delta)
\end{equation}
The first equivalent partner (\ref{t}), looks like the aJC Hamiltonian given in 
Ref.~\cite{negro24}. The second equivalent partner (\ref{s}) is the same as the initial JC Hamiltonian except for the change $\delta \to -\delta$ (see Fig.~\ref{f2}). In other words, if the detuning coefficient $\delta$  change the sign to $-\delta$, we obtain a partner Hamiltonian, $H_{\rm aJC}(\delta)
= H_{\rm JC}(-\delta)$ \cite{lara05,solano03}. This interpretation seems  the most adequate and hereafter we will refer to this choice with the notation $\widetilde{H}_{\rm JC}^{(0)}$ of (\ref{s}).

The energies and eigenvalues of this partner Hamiltonian $\widetilde H_{\rm JC}$ are
\begin{equation}\begin{array}{ll}
\widetilde \varepsilon_n^\pm = n \pm \sqrt{\delta^2 +n \lambda^2},\qquad
&\widetilde\Psi_n^\pm = \left(\begin{array}{c}
(-\delta\pm \sqrt{\delta^2 +n \lambda^2})\,\psi_{n}
\\[1ex]
\sqrt{n}\lambda\, \psi_{n-1}
\end{array}\right)
\\[4.5ex]
\widetilde \varepsilon_0 = \delta,\qquad
&\widetilde\Psi_0 = \left(\begin{array}{c}
\psi_0
\\[1ex]
 0
\end{array}\right)
\end{array}
\end{equation}
Note that the spectrum has the same energies as $H_{\rm JC}$ except for the fundamental states (see Fig.~\ref{f2}):
\begin{equation}
 \varepsilon_0 = -\delta,\qquad  \tilde \varepsilon_0 = \delta
\end{equation} 
The intertwining operator $L$ transforms eigenfunctions of $H_{\rm JC}$ into
eigenfunctions of $ H_{\rm aJC}$ with the same eigenvalue, while the 
adjoint $L^+$ realizes the correspondence in the opposite way,
\begin{equation}
L\, \Psi_n^\pm \ \to \ \widetilde\Psi_n^\pm,\qquad
L^+ \,\widetilde\Psi_n^\pm \ \to \ \Psi_n^\pm,\qquad
\varepsilon_n^\pm = \widetilde \varepsilon_n^\pm\,, \quad n=1,2,\dots
\end{equation}
The action of $L$ and $L^+$ on the respective ground states of $H_{\rm JC}$ and
$\tilde H_{\rm JC}$ is null:
\begin{equation}\label{l00}
L \Psi^-_0 =0\qquad
L^+ \widetilde\Psi^+_0 =0  
\end{equation}
This implies that both Hamiltonians are isospectral, except for the ground states (see Fig.~\ref{f2}).
The connection is carried out by the matrix intertwining operators $L$ and $L^+$, 
where
\begin{equation}
L^+= \left(\begin{array}{cc}
-a^- & 0
\\
0 & a^+
\end{array}\right)
\end{equation}
The product $L^+ L$ gives a symmetry of $H_{\rm JC}$; in fact it can be identified to
the excitation number $N_e$ \cite{gerry04}:
\begin{equation}
N_e:=L^+L= \left(\begin{array}{cc}
a^-a^+ & 0
\\
0 & a^+a^-
\end{array}\right)
\end{equation}
The invariant spaces $V_n$ of (\ref{vn}) correspond to the eigenspaces of 
$L^+L$ with eigenvalue $n$.

The product $L L^+$ will be a symmetry of $\widetilde{H}_{\rm JC}^{(0)}=H_{\rm aJC}$  which can be identified to
the excitation number of the anti-Jaynes-Cummings Hamiltonian,
\begin{equation}
LL^+= \left(\begin{array}{cc}
a^+a^- & 0
\\
0 & a^-a^+
\end{array}\right)
\end{equation}


If we compute the (anti) partner of $ H_{\rm aJC}$, we will recuperate 
$ H_{\rm JC}$. Therefore, the anti-anti-JC Hamiltonian is again $H_{\rm JC}$.

\sect{Second example: Detuned partner JC Hamiltonians}

\subsection{JC Hamiltonian partners by two nonphysical states}

{\bf First option: $\Phi_0^+$ and $\Phi_{-1}^-$.}

In this case, we choose the seed states $\Psi_{\rm s1}$ and $\Psi_{\rm s2}$  as follows
\begin{equation}
\Psi_{\rm s1}\ \to \ \Phi_0^+,\qquad 
\Psi_{\rm s2}\ \to \ \Phi_{-1}^-
\end{equation}
In other words, one of them is the nonphysical ground state $\Phi^+_0$ and the other is the nonphysical excited state $\Phi_{-1}^-$, respectively. Then, we obtain  a triangular intertwining operator 
\begin{equation}\label{int}
L_1= \left(\begin{array}{cc}
a^+ & K_1
\\
0 & a^+
\end{array}\right)\,,\qquad 
K_1 = \frac{-\delta+\sqrt{\delta^2 - \lambda^2}}{\lambda}
\end{equation}
The triangular form of the intertwining operator is a consequence of the seed functions which determine the SUSY transformation. Other choices of seed functions lead us to differential intertwining operators which do not admit an operator expression, so they were discarded.  The partner Hamiltonian will be
\begin{equation}
\widetilde H_{\rm JC}=\left(
\begin{array}{cc}  
a^- a^+ -1 + \sqrt{\delta^2 - \lambda^2} & \lambda a^-
\\[1ex]
\lambda a^+ & a^+a^- - 1 -\sqrt{\delta^2 - \lambda^2}
\end{array}\right)
\end{equation}
This partner Hamiltonian is called  detuned JC Hamiltonian, in the sense that the change in $\widetilde H_{\rm JC}$ with respect to $H_{\rm JC}$ is that the detuning parameter 
$\delta$ has changed to $\sqrt{\delta^2 - \lambda^2}$, besides the subtraction   of a global unit constant. This is possible if the parameters satisfy $\delta^2 \geq \lambda^2$.
We can write this as follows
\begin{equation}
H_{\rm JC}(\delta):= H_{\rm JC}\,,\qquad
\widetilde H_{\rm JC}^{(1)}:= H_{\rm JC}(\sqrt{\delta^2 - \lambda^2})-1
\end{equation}
In this case we may interpret this transformation as a relation between two JC systems with the same radiation but
different atom frequency between the two levels. The eigenvalues and eigenfunctions
of this partner Hamiltonian are as follows
\begin{equation}\begin{array}{ll}
\widetilde\varepsilon_n^\pm = (n-1) \pm \sqrt{\delta^2 +(n-1) \lambda^2},\qquad
&\widetilde\Psi_n^\pm = \left(\begin{array}{c}
(\sqrt{\delta^2 - \lambda^2}\pm \sqrt{\delta^2 +(n-1) \lambda^2})\,\psi_{n-1}
\\[1ex]
\sqrt{n}\lambda\, \psi_n
\end{array}\right)
\\[4.5ex]
\widetilde \varepsilon_0 = -1 - \sqrt{\delta^2 - \lambda^2},\qquad
&\widetilde \Psi_0 = \left(\begin{array}{c}
0
\\[1ex]
 \psi_0
\end{array}\right)
\end{array}
\end{equation}

\begin{figure}[h!]
\begin{subfigure}{0.5\textwidth}
\begin{center}\hskip-0.4cm 
\includegraphics[width=0.7\textwidth]{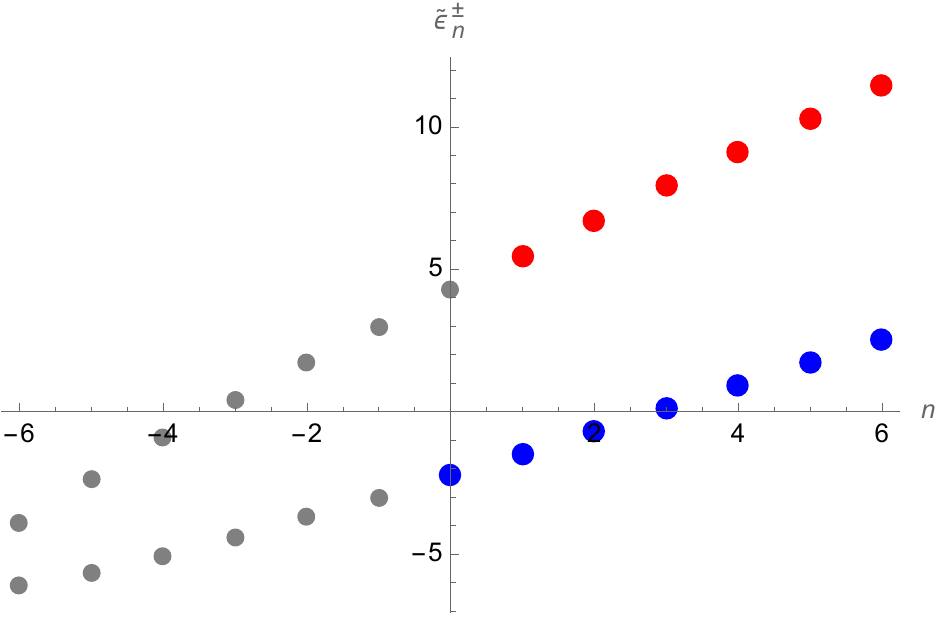}
\caption{\small  Spectrum of $\widetilde H_{\rm JC}^{(-1)}$ }
\end{center}
\end{subfigure}
\begin{subfigure}{0.5\textwidth}
\begin{center}\hskip-0.4cm 
\includegraphics[width=0.7\textwidth]{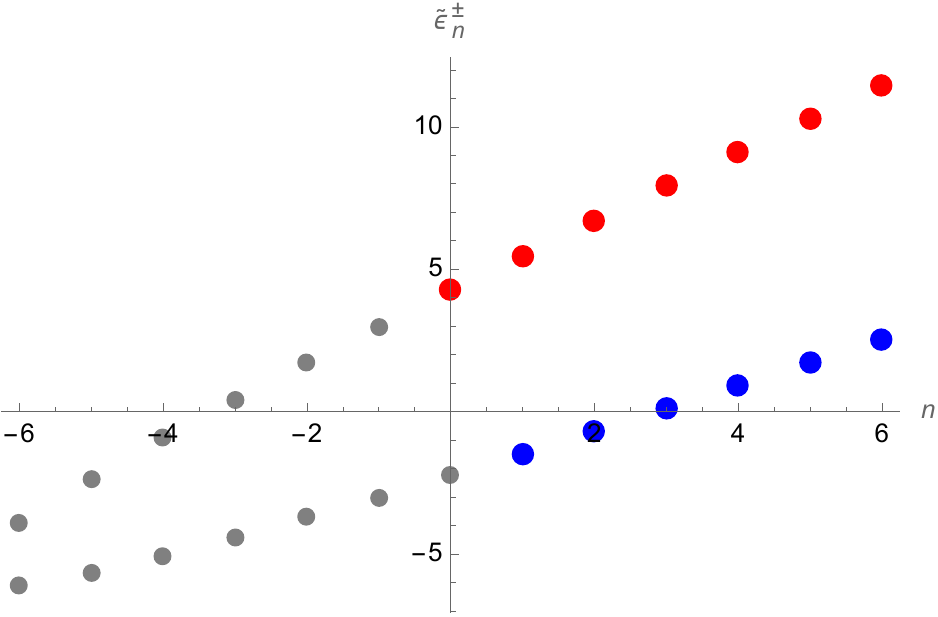}
\caption{\small Spectrum of $\widetilde H_{\rm aJC}^{(-1)}$ }
\end{center}
\end{subfigure}
\\[0.1ex]
\begin{subfigure}{0.5\textwidth}
\quad \begin{center}
\hskip-0.1cm 
\includegraphics[width=0.7\textwidth]{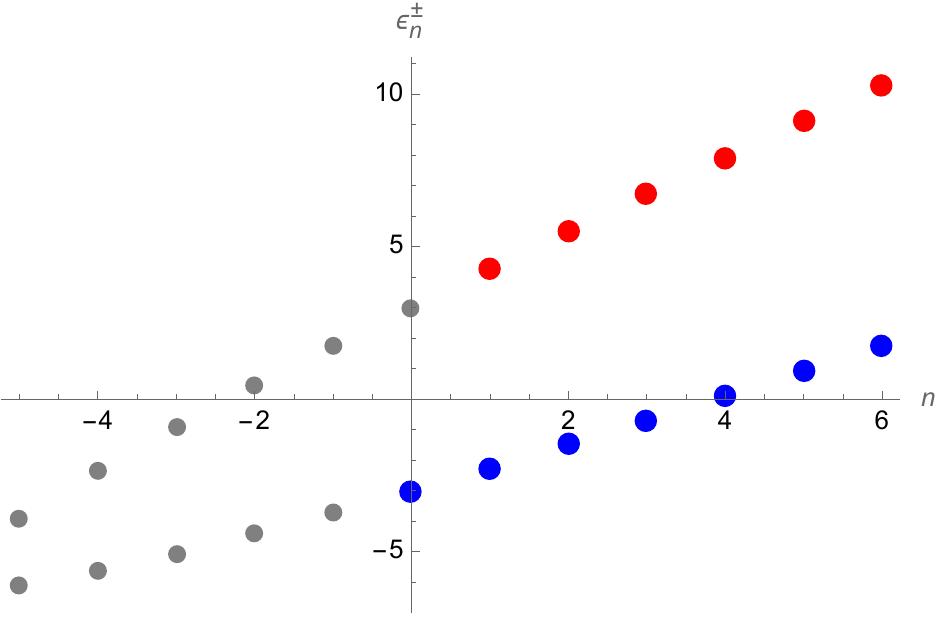}
\caption{\small  Spectrum of $\widetilde H_{\rm JC}^{(0)}$ }
\end{center}
\end{subfigure}
\begin{subfigure}{0.5\textwidth}
\begin{center} 
\hskip-0.1cm 
\includegraphics[width=0.7\textwidth]{figure2.pdf}
\caption{\small Spectrum of $\widetilde H_{\rm aJC}^{(0)}$ }
\end{center}
\end{subfigure}
\\[0.1ex]
\begin{subfigure}{0.5\textwidth}
\begin{center} 
\hskip0.2cm
\includegraphics[width=0.7\textwidth]{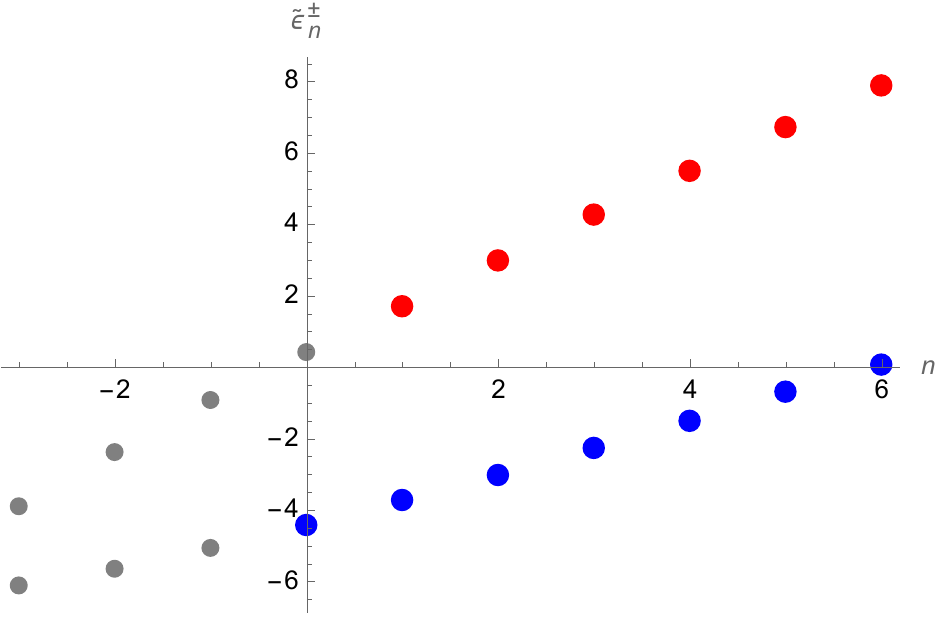}
\caption{\small  Spectrum of $\widetilde H_{\rm JC}^{(1)}$ }
\end{center}
\end{subfigure}
\begin{subfigure}{0.5\textwidth}
\begin{center}
\hskip0.2cm
\includegraphics[width=0.7\textwidth]{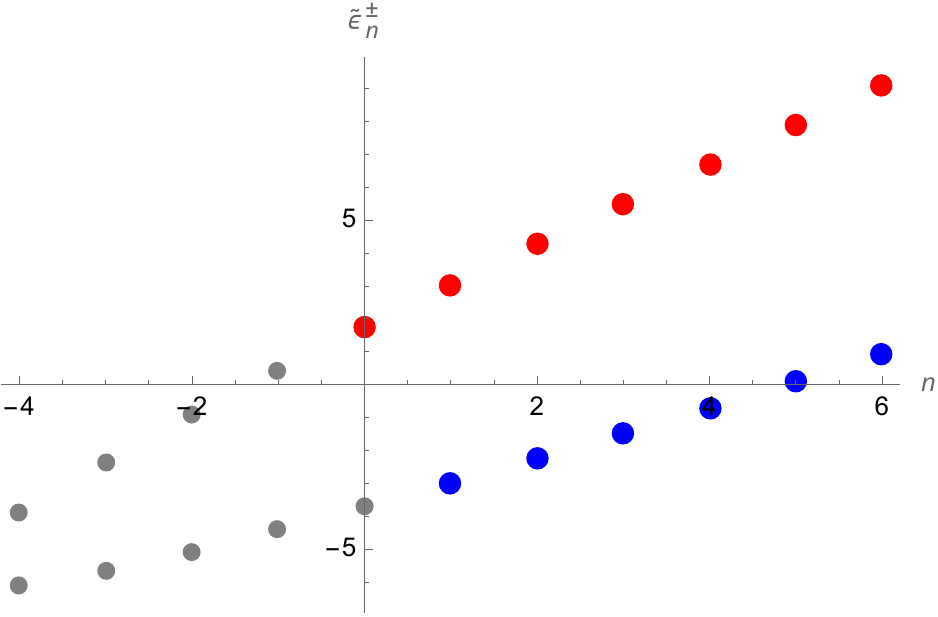}
\caption{\small Spectrum of $\widetilde H_{\rm aJC}^{(1)}$}
\end{center}
\end{subfigure}
\\[0.1ex]
\begin{subfigure}{0.5\textwidth}
\begin{center} 
\hskip0.2cm
\includegraphics[width=0.7\textwidth]{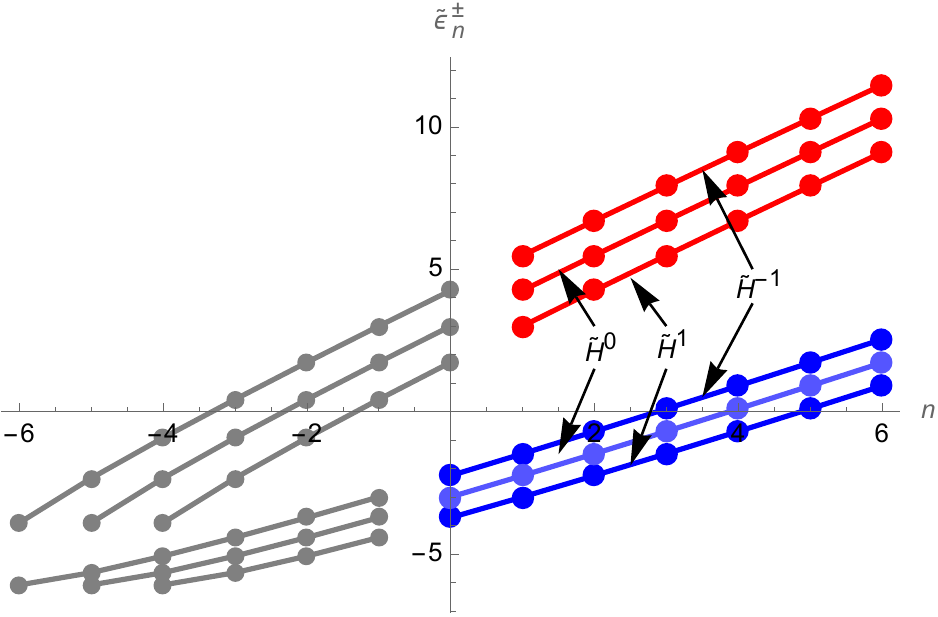}
\caption{\small  Superposition of $\widetilde H_{\rm JC}^{(-1)}$,
$\widetilde H_{\rm JC}^{(0)}$, $\widetilde H_{\rm JC}^{(1)}$ }
\end{center}
\end{subfigure}
\begin{subfigure}{0.5\textwidth}
\begin{center}
\hskip0.2cm
\includegraphics[width=0.7\textwidth]{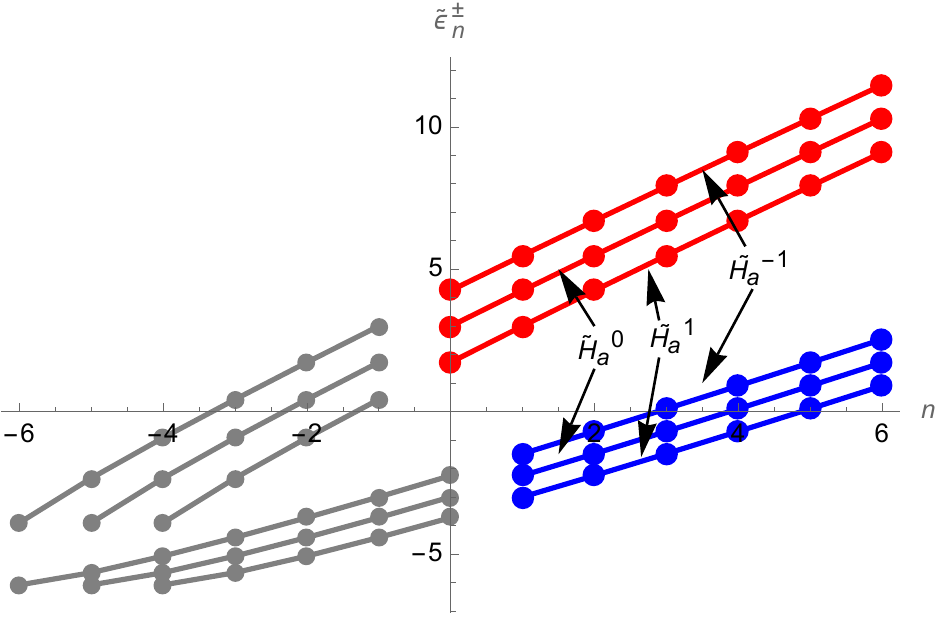}
\caption{\small Superposition of $\widetilde H_{\rm aJC}^{(-1)}$,
$\widetilde H_{\rm aJC}^{(0)}$, $\widetilde H_{\rm aJC}^{(1)}$ }
\end{center}
\end{subfigure}

\caption{\small Left column (for $\delta=3$, $\lambda=1.25$):  spectrum of  JC Hamiltonians  $\tilde H_{\rm JC}^{(-1)}$, $H_{\rm JC}^{(0)}$,
$\tilde H_{\rm JC}^{(1)}$. Points in red ($\tilde\varepsilon_n^{+}$) and  in blue ($\tilde\varepsilon_n^{-}$)  are for the physical spectrum, while gray is for nonphysical states. Right column:  spectrum for the aJC Hamiltonians $\tilde H_{\rm aJC}^{(-1)}$, $\tilde H_{\rm aJC}^{(0)}$,
$\tilde H_{\rm aJC}^{(1)}$. The values of the spectra for these three Hamiltonians are decreasing:
$\tilde\varepsilon_n^{(-1)}>\tilde\varepsilon_n^{(0)} >\tilde\varepsilon_n^{(1)}$.
\label{f3} }
\end{figure}

 The connection of the states and energies are as follows
\begin{equation}\label{tilde}
L_1: \Psi_n^\pm \ \to \widetilde \Psi_{n+1}^\pm,\qquad 
L_1^+: \widetilde \Psi_{n+1}^\pm  \ \to  \ \Psi_n^\pm\,,
\qquad 
\varepsilon_n^\pm = \widetilde \varepsilon_{n+1}^\pm,\qquad n=1,2,\dots
\end{equation}
But the action on the nonphysical states is null:
\begin{equation}\label{l11}
L_1 \Phi_0^+ =0,\qquad L_1 \Phi_{-1}^- =0\qquad
\end{equation}
This means that both Hamiltonians are isospectral, except for two levels: One ground and one excited states.
 To be more precise, 
$\varepsilon_0 = \widetilde \varepsilon^{ \,-}_1$, but $\widetilde \varepsilon_0^+$ and $\widetilde \varepsilon^{ \,+}_{1}$ are two new levels of the partner Hamiltonian which were not present in the initial $H_{\rm JC}$. The other energy levels are in one to one correspondence: $\varepsilon_n^\pm = \widetilde \varepsilon_{n+1}^\pm$, for $n=1,2,\dots$ as mentioned in (\ref{tilde}).
The connection between the states of both Hamiltonians  is carried out by the matrix intertwining operator $L_1$. In conclusion, we see that a smaller positive detuning $\delta$ leads to increase the spectrum with two more energy eigenvalues which are at the bottom of the spectrum, see Fig.~\ref{f3}e.

By using this intertwining operator we can find a symmetry, $S_1$ of the Hamiltonian:
\begin{equation}
S_1 = L^+_1 \, L_1 = 
\left(
\begin{array}{cc}
a^-a^+ &  a^- K_1
\\
a^+ K_1 & a^-a^+ + K_1^2 
\end{array}
\right)\,,\qquad K_1= \frac{-\delta+\sqrt{\delta^2 - \lambda^2}}{\lambda}
\end{equation}
This symmetry is a linear combination of the previous symmetries:
\begin{equation}
S_1 = N_e + \frac{K_1}{\lambda}(H_{\rm JC} -N_e -\delta)
\end{equation}

%

{\bf Second option: $\Phi^+_0$ and $\Phi_{-1}^+$.}

In this case, we choose the nonphysical seed states $\Psi_{\rm s1}$ and $\Psi_{\rm s2}$   as follows
\begin{equation}
\Psi_{\rm s1}\ \to \ \Phi_0^+,\qquad 
\Psi_{\rm s2}\ \to \ \Phi_{-1}^+
\end{equation}
In other words, one of them is the nonphysical ground state $\Phi_0^+$ and the other is the nonphysical excited state, $\Phi_{-1}^+$ respectively. Then, we obtain  the intertwining operator 
\begin{equation}\label{int}
L_2= \left(\begin{array}{cc}
a^+ & -K_2^+
\\
0 & a^+
\end{array}\right)\,,\qquad 
K_2^+ = \frac{\delta+\sqrt{\delta^2 - \lambda^2}}{\lambda}
\end{equation}
The partner Hamiltonian will be
\begin{equation}\label{np2}
\widetilde H_{\rm aJC}^{(1)}:=\widetilde H_{\rm JC}=\left(
\begin{array}{cc}  
a^- a^+ -1 - \sqrt{\delta^2 - \lambda^2} & \lambda a^-
\\[1ex]
\lambda a^+ & a^+a^- - 1 +\sqrt{\delta^2 - \lambda^2}
\end{array}\right)
\end{equation}
This partner Hamiltonian is a  detuned anti-JC Hamiltonian. If we make an equivalence transformation
\begin{equation}
\widetilde H_{\rm JC} \ \to \ \widetilde H_{\rm aJC}=T \widetilde H_{\rm JC} T^{-1}\,,\quad {\rm where}
\quad \ T=
\left(\begin{array}{cc}
0 & 1
\\[1.5ex]
-1 & 0
\end{array}\right)
\end{equation}
we have 
\begin{equation}
\widetilde H_{\rm aJC}=\left(
\begin{array}{cc}  
a^+ a^- -1 + \sqrt{\delta^2 - \lambda^2} & -\lambda a^+
\\[1ex]
-\lambda a^- & a^-a^+ - 1 -\sqrt{\delta^2 - \lambda^2}
\end{array}\right)
\end{equation}
Therefore, $\widetilde H_{\rm aJC}$ has the form of an anti-JC Hamiltonian like (\ref{ajc}) where the detuning parameter 
$\delta$ has changed to $\sqrt{\delta^2 - \lambda^2}$, besides a global constant. 

In this case, we have also the following interpretation: We are relating two JC systems with of opposite  detuning sign as seen in (\ref{np2}). The eigenvalues and eigenfunctions
of this partner Hamiltonian are as follows
\begin{equation}\begin{array}{ll}
\widetilde\varepsilon_n^\pm = (n-1) \pm \sqrt{\delta^2 +(n-1) \lambda^2},\qquad
&\widetilde\Psi_n^\pm = \left(\begin{array}{c}
(-\sqrt{\delta^2 - \lambda^2}\pm \sqrt{\delta^2 +(n-1) \lambda^2})\,\psi_{n-1}
\\[1ex]
\sqrt{n}\lambda\, \psi_n
\end{array}\right)
\\[4.5ex]
\widetilde \varepsilon_0^+ = -1 + \sqrt{\delta^2 - \lambda^2},\qquad
&\widetilde \Psi_0 = \left(\begin{array}{c}
0
\\[1ex]
 \psi_0
\end{array}\right)
\end{array}
\end{equation}


 The connection of the states and energies are as follows
\begin{equation}
L_2: \Psi_n^\pm \ \to \widetilde \Psi_{n+1}^\pm,\qquad 
\varepsilon_n^\pm = \widetilde \varepsilon_{n+1}^\pm
\end{equation}
But the action on the nonphysical states is null:
\begin{equation}
L_2 \Phi_0^+ =0,\qquad L_2 \Phi_{-1}^+ =0\qquad
\end{equation}
This means that both Hamiltonians are isospectral, except for two lower energy levels:
$\widetilde \varepsilon^+_0$ and $\widetilde \varepsilon_{-1}^+$, see Fig.~\ref{f3}f.

The connection between the states of both Hamiltonians  is carried out by the matrix intertwining operator $L_2$. 
In this case, the symmetry is the following linear combination of the previous ones:
\begin{equation}
S_2 = L_2^+L_2= N_e - \frac{K_2^+}{\lambda}(H_{\rm JC} -N_e -\delta)
\end{equation}

\subsection{JC Hamiltonian partners by two physical states}

{\bf First option: $\Psi_0^-$ and $ \Psi_1^+$.}

In this case, we choose the seed states $\Psi_{\rm s1}$ and $\Psi_{\rm s2}$  as follows
\begin{equation}
\Psi_{\rm s1}\ \to \ \Psi_0,\qquad 
\Psi_{\rm s2}\ \to \ \Psi_1^+
\end{equation}
So, one of them is the  physical ground state $\Psi_0$ an the other is the physical first excited state, $\Psi_1^+$ respectively. Then, we obtain  the intertwining operator 
\begin{equation}\label{int}
L_3= \left(\begin{array}{cc}
a^- & 0
\\
K_3 & a^-
\end{array}\right)\,,\qquad 
K_3= \frac{\delta- \sqrt{\delta^2 + \lambda^2}}{\lambda}
\end{equation}
The partner Hamiltonian will be
\begin{equation}
\widetilde H_{\rm JC}^{(-1)}:=\widetilde H_{JC}=\left(
\begin{array}{cc}  
a^- a^+ +1 + \sqrt{\delta^2 + \lambda^2} & \lambda a^-
\\[1ex]
\lambda a^+ & a^+a^- + 1 -\sqrt{\delta^2 + \lambda^2}
\end{array}\right)
\end{equation}
The eigenvalues and eigenfunctions are
\begin{equation}\begin{array}{ll}
\widetilde\varepsilon_n^\pm = (n+1) \pm \sqrt{\delta^2 + (n+1) \lambda^2},\qquad
&\widetilde\Psi_n^\pm = \left(\begin{array}{c}
(\sqrt{\delta^2 + \lambda^2}\pm \sqrt{\delta^2 + (n+1) \lambda^2})\,\psi_{n-1}
\\[1ex]
\sqrt{n}\lambda\, \psi_n
\end{array}\right)
\\[4.5ex]
\widetilde \varepsilon_0 = 1 - \sqrt{\delta^2 + \lambda^2},\qquad
&\widetilde \Psi_0 = \left(\begin{array}{c}
0
\\[1ex]
 \psi_0
\end{array}\right)
\end{array}
\end{equation}
 
In this case, the new detuning parameter is bigger: $\delta \to \sqrt{\delta^2 + \lambda^2}$. 
The connection of the states and energies are as follows
\begin{equation}
L_3 \Psi_n^\pm \ \to \widetilde \Psi_{n-1}^\pm,\qquad 
\varepsilon_n^\pm = \widetilde \varepsilon_{n-1}^\pm
\end{equation}
But the action on the physical states is null:
\begin{equation}
L_3 \Psi_0 =0,\qquad L_3 \Psi_1^+ =0\qquad
\end{equation}
This implies that this partner Hamiltonian has lost two energy levels (see Fig.~\ref{f3}a) corresponding to these annihilated states.

By using this intertwining operator we can find a symmetry, $S_3$ of the Hamiltonian:
\begin{equation}
S_3 = L_3^+\, L_3 = 
\left(
\begin{array}{cc}
a^+a^- + K_3^2&  a^- K_3
\\
a^+ K_3 & a^+a^-
\end{array}
\right)\,,\qquad K_3= \frac{\delta-\sqrt{\delta^2 + \lambda^2}}{\lambda}
\end{equation}
This symmetry is a linear combination of the previous symmetries:
\begin{equation}
S_3 =L_3^+\, L_3 =  N_e + \frac{K_3}{\lambda}(H_{\rm JC} -N_e +\delta)
\end{equation}



{\bf Second option: $\Psi_0$ and $ \Psi_1^-$.}

In this case, we choose the seed states $\Psi_{\rm s1}$ and $\Psi_{\rm s2}$  as follows
\begin{equation}
\Psi_{\rm s1}\ \to \ \Psi_0,\qquad 
\Psi_{\rm s2}\ \to \ \Psi_1^-
\end{equation}
So, one of then is the  physical ground state $\Psi_0$ an the other is the physical first excited state, $\Psi_1^-$ respectively. Then, we obtain  the intertwining operator 
\begin{equation}\label{int}
L_4= \left(\begin{array}{cc}
a^- & 0
\\
K_4 & a^-
\end{array}\right)\,,\qquad 
K_4= \frac{\delta+\sqrt{\delta^2 + \lambda^2}}{\lambda}
\end{equation}
The partner Hamiltonian will be
\begin{equation}
\widetilde H_{\rm aJC}^{(-1)}:=\widetilde H_{\rm JC}=\left(
\begin{array}{cc}  
a^- a^+ +1 - \sqrt{\delta^2 + \lambda^2} & \lambda a^-
\\[1ex]
\lambda a^+ & a^+a^- + 1 +\sqrt{\delta^2 + \lambda^2}
\end{array}\right)
\end{equation}

This parter Hamiltonian  has changed $\delta$ by $- \sqrt{\delta^2 + \lambda^2}$, so it is of aJC type. The eigenvalues and eigenfunctions are
\begin{equation}\begin{array}{ll}
\widetilde\varepsilon_n^\pm = (n+1) \pm \sqrt{\delta^2 +(n+1) \lambda^2},\qquad
&\widetilde\Psi_n^{\,\pm} = \left(\begin{array}{c}
(-\sqrt{\delta^2 - \lambda^2}\pm \sqrt{\delta^2 + (n+1) \lambda^2})\,\psi_{n-1}
\\[1ex]
\sqrt{n}\lambda\, \psi_n
\end{array}\right)
\\[4.5ex]
\widetilde \varepsilon_0 = 1 + \sqrt{\delta^2 + \lambda^2},\qquad
&\widetilde \Psi_0 = \left(\begin{array}{c}
0
\\[1ex]
 \psi_0
\end{array}\right)
\end{array}
\end{equation}
%

The connection of the states and energies are as follows
\begin{equation}
L_4\Psi_n^\pm \ \to \widetilde \Psi_{n-1}^\pm,\qquad 
\varepsilon_n^\pm = \widetilde \varepsilon_{n-1}^\pm
\end{equation}
But the action on the physical seed states is null:
\begin{equation}
L_4\Psi_0 =0\,,\qquad L_4 \Psi_1^- =0\qquad
\end{equation}
This implies that the partner Hamiltonian has also lost two energy levels (see Fig.~\ref{f3}b) corresponding to these annihilated states.

By using this intertwining operator we can find a symmetry, $S_4$ of the Hamiltonian:
\begin{equation}
S_4 = L_4^+\, L_4 = 
\left(
\begin{array}{cc}
a^+a^- + K_4^2&  a^- K_4
\\
a^+ K_4 & a^+a^-
\end{array}
\right)\,,\qquad K_4= \frac{\delta+\sqrt{\delta^2 + \lambda^2}}{\lambda}
\end{equation}

This symmetry is a linear combination of the previous symmetries:
\begin{equation}
S_4 =L_4^+\, L_4=  N_e + \frac{K_4}{\lambda}(H_{\rm JC} -N_e +\delta)
\end{equation}

\subsection{Summary of JC Hamiltonians  SUSY partners}



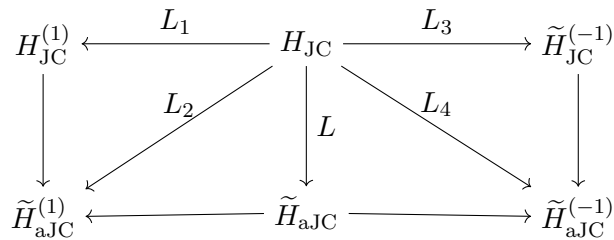
\begin{figure}[h!]
\begin{center}
\begin{tikzpicture}[trafo/.style={midway
}]

  \def\hd{2.5}\def\vd{1.5}\def\vdd{1.6}

  \node (hm1) at (0,-0.5) {$H_{\rm JC}^{(1)}$};
  \node[right=\hd of hm1] (h0) {$H_{\rm JC}$};
  \node[right=\hd of h0] (h1) {$\widetilde 
H_{\rm JC}^{(-1)}$};

  \node[below=\vd of hm1] (hhm1) {$\widetilde H_{\rm aJC}^{(1)}$};
  \node[below=\vdd of h0] (hh0) {$\widetilde H_{\rm aJC}$};
  \node[below=\vd of h1] (hh1) {$\widetilde 
H_{\rm aJC}^{(-1)}$};

  \draw[<-] (hm1) -- (h0) node[trafo,above] {$L_1$};
  \draw[->] (h0) -- (h1) node[trafo,above] {$L_3$};
  \draw[->] (h1) -- (hh1);
  \draw[->] (h0) -- (hhm1) node[trafo,above] {$L_2$};
  \draw[->] (h0) -- (hh1) node[trafo,above] {$L_4$};

  \draw[->] (h0) -- (hh0) node[trafo,right] {$L$};
  ;
  \draw[->] (hm1) -- (hhm1) 
  ;

  \draw[<-] (hhm1) -- (hh0) 
  ;
  \draw[->] (hh0) -- (hh1) 
  ;

\end{tikzpicture}
\medskip
\caption{This diagram shows all the partner Hamiltonians of $H_{\rm JC}(\delta)$ obtained along Sections 4 and 5. The intertwining operators are $L$, $L_1,L_2,L_3,L_4$.}
\label{fig7}
\end{center}
\end{figure}

As we have seen above, the SUSY transformations constructed in this work  act on a JC Hamiltonian characterized by a detuning parameter $\delta$ giving rise to a kind of anti-JC Hamiltonian and four types of partner Hamiltonians with different detuning parameters, as shown in the diagram of Fig.~\ref{fig7}. 

\bigskip

\bigskip

We can obtain new Hamiltonians by consecutive SUSY transformations, in this way we arrive at a sequence of JC Hamiltonians starting from $H_{\rm JC}(\delta)$, as follows,
\begin{equation}
(i)\qquad \leftarrow\ H_{\rm JC}^{(-n)}\  \leftarrow \dots \ H_{\rm JC}^{(-1)}\ \leftarrow \ H_{\rm JC}(\delta) \ \to \ H_{\rm JC}^{(1)}
\dots \ \to \ \ H_{\rm JC}^{(n)}\ \to
\end{equation}
where
\begin{equation}
H_{\rm JC}^{(n)}=H_{\rm JC}(\sqrt{\delta^2-n\lambda^2})-n,\qquad
H_{\rm JC}^{(-n)}=H_{\rm JC}(\sqrt{\delta^2+n\lambda^2})+n
\end{equation}
From $H_{\rm JC}(\delta)$ we obtained an anti-JC Hamiltonian, 
$H_{\rm aJC}$, which was equivalent to $H_{\rm JC}(-\delta)$. Applying sucessive
SUSY transformation to $H_{\rm aJC}$ we get a second sequence 
\begin{equation}
(ii)\qquad
\leftarrow\ H_{\rm aJC}^{(-n)}\  \leftarrow \dots \ H_{\rm aJC}^{(-1)}\ \leftarrow \ H_{\rm aJC}(\delta) \ \to \ H_{\rm aJC}^{(1)}
\dots \ \to \ \ H_{\rm aJC}^{(n)}\ \to
\end{equation}
where
\begin{equation}
H_{\rm aJC}^{(n)}\approx H_{\rm JC}(-\sqrt{\delta^2-n\lambda^2})-n,\qquad
H_{\rm aJC}^{(-n)}\approx H_{\rm JC}(-\sqrt{\delta^2+n\lambda^2})+n
\end{equation}

%
%
%


\sect{The resonant hierarchy of JC Hamiltonians}

\begin{figure}[h!]
\begin{center}
\includegraphics[width=0.50\textwidth]{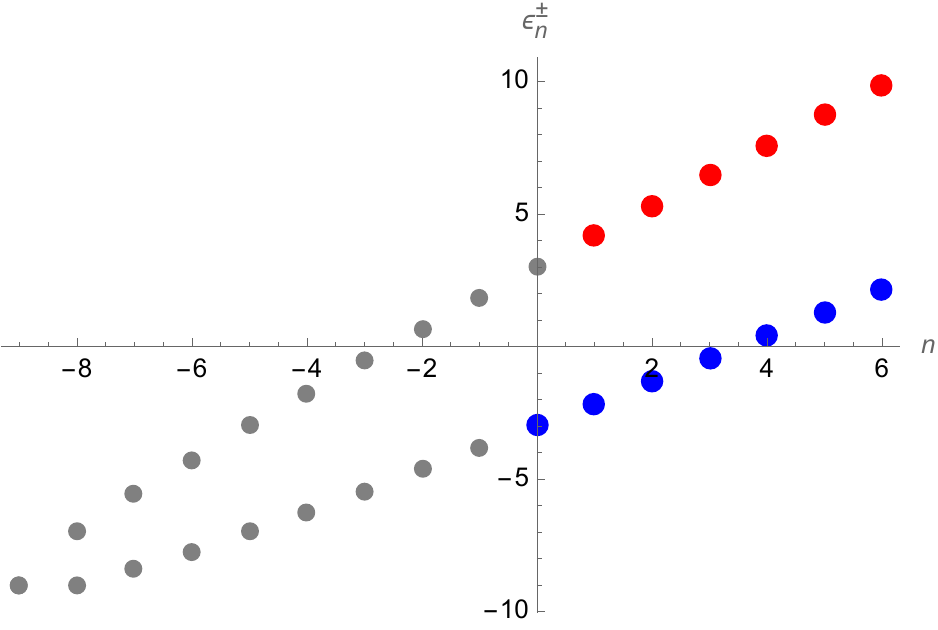}
\end{center}
\caption{\small Eigenvalues of JC Hamiltonian of the resonant type, for $\delta=3$, $\lambda=1$. Blue and red points are for the (infinite) physical spectrum, while
gray for a (finite) nonphysical spectrum.  It is characterized by a common energy eigenvalue for both initial ground states $\Psi_0^+$ and $\Psi_0^-$, which corresponds to the left gray point in the figure.
\label{f8}
}
\end{figure}

In the particular case where $\delta^2 = k \lambda^2$, for $k$ a positive integer, after $k$ SUSY transformations, we can
obtain a partner Hamiltonian  $ H_{\rm JC}(\delta' =0)$, which is a resonant
JC Hamiltonian. This type of Hamiltonians, which will be denoted
$H_{\rm JC}^k$,  constitute a ``resonant hierarchy" and they have a very special form:
\begin{equation}
 H_{\rm JC}^k=
\left(
\begin{array}{cc}  
a^- a^+ +\lambda\sqrt{k} &\lambda a^-
\\[1ex]
\lambda a^+ & a^+a^- -\lambda\sqrt{k}
\end{array}\right) \ +k I \,, \qquad k= 0,1,2\dots
\end{equation}
The spectrum of one example in this class, including the unphysical part,  is given in Fig.~\ref{f8}, where it is shown that the two sequences of physical and nonphysical spectrum end in one point (not two as it is the generic case). These Hamiltonians are intertwined to the resonant Hamiltonian $H_{\rm JC}^0$
by means of the operators having the form
\begin{equation}\label{int2}
L_k= \left(\begin{array}{cc}
-a^+ &  \sqrt{k}- \sqrt{k-1} 
\\
0 & -a^+
\end{array}\right) 
\end{equation}
and the intertwining relations
\begin{equation}\label{int3}
L_k  H_{\rm JC}^k =  H_{\rm JC}^{k-1} L_k\,, \qquad  k= 1,2,\dots
\end{equation}

The Hamiltonians $ H_{\rm JC}^k$ have $2k$ states less than the resonant one (each transformation looses two states). We can express this class of Hamiltonians in the form
\begin{equation}
 H_{\rm JC}^k=
\left(
\begin{array}{cc}  
a^- a^+  &0
\\[1ex]
0 & a^+a^- 
\end{array}\right) 
+ 
\lambda\,\left(\begin{array}{cc}  
\sqrt{k} & a^-
\\[1ex]
 a^+ &  -\sqrt{k}
\end{array}\right)
= N_e + \lambda H_{\rm HO}^k
\end{equation}
where $N_e$ is excitation number  operator, while 
$H_{\rm HO}^k$ is a first order Hamiltonian associated to the harmonic oscillator. 
We find that
\begin{equation}
\big(H_{\rm HO}^k\big)^2 = N_e + k
\end{equation}
The Hamiltonians $H_{\rm HO}^k$ constitute a hierarchy of shape invariant Dirac Hamiltonians which was considered in Ref.~\cite{kuru21}. Needless to say,  there is also an anti-JC Hamiltonian resonant hierarchy with $\delta= - \sqrt{n} \lambda$.

\sect{Conclusions}

In this work, we have shown how SUSY transformations can be applied to  JC Hamiltonians. These JC Hamiltonians are characterized by  a pair of parameters:
The interacting coefficient $\lambda$ and the detuning $\delta$; what we have seen is that the Hamiltonian $H_{\rm JC}(\delta,\lambda)$ can be transformed
into other two $\widetilde H^{(\pm1)}_{\rm JC}=H_{\rm JC}(\delta'_\pm,\lambda)\pm1$, with  different detuning parameters
$\delta'_\pm= \sqrt{\delta^2 \pm \lambda^2}$. These two Hamiltonians have the same spectra except for two of the lowest points of the spectrum.  

We find that there is the possibility to change  the sign of the detuning by means of a SUSY transformation, so that we can transform  $H_{\rm JC}(\delta,\lambda)$ into $H_{\rm JC}(-\delta,\lambda)$
or into $H_{\rm JC}(\delta',\lambda)$ with $\delta'= -\sqrt{\delta^2 \pm \lambda^2})$. In this way, we find  two sequences of JC Hamiltonians, the so called JC and anti-JC hierarchies corresponding to positive or negative $\delta$ values, connected by special SUSY transformations. 
Note that the transformations here involved were constructed by means of appropriate physical as well as nonphysical eigenstates called seed functions.

Thus, any sequence or hierarchy is connected by SUSY transformations, such that any two of their consecutive Hamiltonians differ in two eigenvalues and they are related by intertwining operators annihilating their corresponding eigenfunctions.
The product of two intertwining operators gives a symmetry of each JC Hamiltonian. In particular, the excitation number is a product of intertwining operators, i.e., it is one of such symmetries. Each Hamiltonian has a particular factorized symmetry.
In fact, many properties of these partner Hamiltonians should have common features due to this relationship. The program to show the similarities and differences related with  expected values of atomic operators (such as $\sigma_z$ for the atomic inversion) or photon operators (such as quadratures), is  postponed to a future work to avoid a too lengthy paper. 

There is a particular hierarchy, where one of the JC Hamiltonian has a vanishing detuning parameter $\delta=0$. This is a very interesting situation where all the intertwining properties have very simplified expressions. This case is similar to previous results on the shape invariance of matrix Hamiltonians of two-dimensional relativistic equations applied in graphene \cite{kuru21}. This is  another example of a quite appealing relation between quantum optics (through JC Hamiltonians)  and condensed matter (of low energy electronic states in graphene ruled by the Dirac-Weyl matrix equation). 

\section*{Data availability statement}
No new data were created or to be shared in this study.

\section*{Acknowledgments}

The authors would like to acknowledge the support of the QCAYLE project, funded by the European Union--NextGenerationEU, PID2020-113406GB-I0 project funded by the MCIN of Spain and the contribution of the European Cooperation in Science and Technology COST Action CA23130.
\c{S}.~Kuru thanks Ankara University and the warm hospitality of the Department of Theoretical Physics of the University of Valladolid, where part of this work has been carried out, and to the support of its GIR of Mathematical Physics.  \.{I}. B.~Ate\c s thanks to TUBITAK BIDEB 2211-A program.


\begin{thebibliography}{99}


\bibitem{jaynes63}  E. T. Jaynes and F. W. Cummings, Comparison of quantum and semiclassical radiation theories with applications to the beam maser, Proc. IEEE,  {\bf51}, (1963) 89-109.
 
\bibitem{rempe87}  G. Rempe and H. Walther, N. Klein, Observation of quantum collapse and revival in a one-atom maser, Phys. Rev. Lett., {\bf58},  (1987) 353.

\bibitem{gerry04}  C. Gerry and P. Knight, Introductory Quantum Optics, Cambridge University Press, Cambridge, (2004).

\bibitem{Rossatto17} D. Z. Rossatto, C. J. Villas-Boas, M. Sanz and E. Solano, Spectral classification of coupling regimes in the quantum Rabi model,
Phys. Rev. A, {\bf 96}, (2017) 013849.

\bibitem{Larson24}
J. Larson, T.  Mavrogordatos, The Jaynes Cummings Model and its Descendants (Second Edition), IOP Publishing, Bristol, UK (2024).

\bibitem{Solano19} P. Forn-Diaz, L. Lamata, E. Rico, J. Kono, E. Solano, Ultrastrong coupling regimes of light-matter interaction, Rev. Mod. Phys., {\bf 91}, (2019) 025005.

\bibitem{Huang20}
J.-F. Huang, J.-Q. Liao, L.-M. Kuang, Ultrastrong Jaynes-Cummings model, Phys. Rev. A, {\bf  101}, (2020) 043835.

\bibitem{Haroche96}
M. Brune, F. Schmidt-Kaler, A. Maali, J. Dreyer, E. Hagley, J. M. Raimond, and S. Haroche, Quantum Rabi oscillation: A direct test of field quantization in a cavity, Phys. Rev. Lett., {\bf 76}, (1996) 1800.

\bibitem{Chong25}
S. Chong, X.  Liu, Y. Han, J. Du, J. Liu, L. Meng, Y. Gao, C. Yang, J. Qi Shen, J.  Zhang, and L. Li, The long-time behavior of collapse-revival effects in a multi-photon Jaynes Cummings model, Journal of the Physical Society of Japan, {\bf 94}, (2025) 024402.

\bibitem{matveev91}   V. B. Matveev and M. A. Salle,  Darboux Transformations and Solitons, Springer, Berlin, (1991).

\bibitem{infeld51}   L. Infeld and T. E. Hull,  The factorization method, Rev. Mod. Phys., {\bf23}, (1951)  21-68.

\bibitem{fernandez10}  D. J. Fern\'andez C., Supersymmetric quantum mechanics,  AIP Conf. Proc.,  {\bf3}, (2010)  1287.

\bibitem{cooper95}  F. Cooper, A. Khare and U. Sukhatme, Supersymmetry and quantum mechanics, Phys. Rep.,  {\bf251}, (1995)  267-385.

\bibitem{junker96} G. Junker,  Supersymmetric Methods in Quantum and Statistical Physics, Springer Verlag, Berlin, (1996).



\bibitem{negro04} B. F. Samsonov, J. Negro, Darboux transformations of the Jaynes-Cummings Hamiltonian, J. Phys. A: Math. Gen., {\bf37}, (2004)  10115-10127.

\bibitem{hussin06}   V. Hussin, \c{S}. Kuru, J. Negro, Generalized Jaynes–Cummings Hamiltonians by shape-invariant hierarchies and their SUSY partners, J. Phys. A: Math. Gen., {\bf39}, (2006)  11301-11311.

\bibitem{kuru21}   D. D. K\i z\i l\i rmak, \c{S}. Kuru, J. Negro,  Dirac-like Hamiltonians associated to Schr\"odinger factorizations, Eur. Phys. J. Plus, {\bf136}, (2021)  668.

\bibitem{negro24} I. A. Bocanegra-Garay, M. Castillo-Celeita
J. Negro, L. M. Nieto and F. J. Gomez-Ruiz,  Exploring supersymmetry: Interchangeability between Jaynes-Cummings and
anti-Jaynes-Cummings models, Phys. Rev., {\bf 6}, (2024) 043218. 

\bibitem{lara24} A. Kafuri, F. H. Maldonado-Villamizar, A. Moroz and B. M. Rodriguez-Lara, A supersymmetry journey from the Jaynes-Cummings to the anisotropic Rabi model, J. Opt. Soc. Am. B,  {\bf 41}, (2024) C82-C90.

\bibitem{lara05}  B. M. Rodr\'{\i}guez-Lara and H. Moya-Cessa, A. B. Klimov, Combining Jaynes-Cummings and anti-Jaynes-Cummings dynamics in a trapped-ion system driven by a laser, Phys. Rev. A, {\bf71}, (2005)  023811.


\bibitem{alhaidari06} A. D. Alhaidari, Supersymmetric Jaynes-Cummings model and its exact solutions, J. Phys.: Math. Gen., {\bf39}, (2006) 15391. 

\bibitem{panahi15}  H. Panahi, L. Jahangiri, Generalized Jaynes-Cummings model and shape invariant potentials: Master function approach, Int J. Theor. Phys., {\bf54}, (2015)  2675-2683.

\bibitem{lara13} B. M. Rodr\'{\i}guez-Lara, F. Soto-Eguibar, A. Zárate, H. M. Moya-Cessa, A classical simulation of nonlinear Jaynes-Cummings and Rabi models in photonic lattices, Opt. Express., {\bf21}, (2013)  12888-12898.

\bibitem{lara21}  F. H. Maldonado-Villamizar, C. A. Gonz\'alez-Gutierrez, L. Villanueva-Vergara and B. M. Rodrguez-Lara, Underlying SUSY in a generalized Jaynes-Cummings model, Sci. Rep., {\bf11}, (2021)  16467.



\bibitem{mubark23}  M. R. Abdel-Allah Mubark, R. M. Farghly, S. H. Mohamed, Supersymmetry as an algebraic approach to the Jaynes-Cumming model with stark shift and Kerr-like medium, AUNJMSR, {\bf52}, (2023)  177-187.

\bibitem{alexio07} A. N. F. Aleixo and A. B. Balantekin, A two-level atom coupled to a two-dimensional supersymmetric and shape-invariant system: models, J. Phys. A: Math. Theor.,  {\bf40}, (2007)  3915.

\bibitem{castanos13}
O. Casta\~nos, Supersymmetry in the Jaynes-Cummings model,
AIP Conf. Proc.,  {\bf 1540}, (2013) 77-83.

\bibitem{solano03} E. Solano, G. S. Agarwal and H. Walther, Strong-driving-assisted multipartite entanglement in cavity QED,  Phys. Rev. Lett., {\bf 90}, (2003) 027903.

\bibitem{ismail25}
M. Kh. Ismail, T. M. El-Shahat, N. Metwally, A. S. F. Obada,
Entangling power for anti-Jaynes-Cummings model,
Optik,
{\bf 325}, (2025) 172244.


\end{thebibliography}
\end{document}